\documentclass[pre,floatfix,pdftex,twocolumn,ltxgrid,superscriptaddress]{revtex4-1}
\usepackage[pdftex]{graphicx} \usepackage{subfigure}
\usepackage{stdclsdv} \usepackage{array} \usepackage{amsmath}

\usepackage{balance} 
\usepackage[english]{babel}
\usepackage{graphicx} \usepackage{subfigure}
\usepackage{stdclsdv} \usepackage{array}
\usepackage{siunitx}
\usepackage{amsmath}
\usepackage{amsfonts}
\usepackage{amssymb}
\usepackage{comment}
\usepackage{tabularx}

\usepackage{bm}
\usepackage{graphicx} 
\usepackage{lastpage}

\usepackage[normalem]{ulem}

\newcommand{\av}[1]{\langle #1 \rangle}

\begin{document}

\title{Fiber networks amplify active stress}

\author{Pierre
  Ronceray}\email{pierre.ronceray@u-psud.fr}~\affiliation{Univ. Paris-Sud;
  CNRS; LPTMS; UMR 8626, Orsay 91405 France.}  \author{Chase
  Broedersz}\email{c.broedersz@lmu.de}~\affiliation{Lewis-Sigler
  Institute for Integrative Genomics and Joseph Henry Laboratories of
  Physics, Princeton University, Princeton, NJ 08544, USA}\affiliation{
  Arnold-Sommerfeld-Center for Theoretical Physics and Center for
  NanoScience, Ludwig-Maximilians-Universit\"at M\"unchen,
  Theresienstrasse 37, D-80333 M\"unchen, Germany.}  \author{Martin
  Lenz}\email{martin.lenz@u-psud.fr}~\affiliation{Univ. Paris-Sud;
  CNRS; LPTMS; UMR 8626, Orsay 91405 France.}

\begin{abstract}
  Large-scale force generation is essential for biological functions
  such as cell motility, embryonic development, and muscle
  contraction. In these processes, forces generated at the molecular level
  by motor proteins are transmitted by disordered fiber networks,
  resulting in large-scale active stresses.
  While these fiber networks are well characterized macroscopically, this stress
  generation by microscopic active units is not well understood.
  Here we
  theoretically study force transmission in these
  networks, and find that local active forces are rectified towards
  isotropic contraction and strongly amplified as fibers
  collectively buckle in the vicinity of the active units. This stress
  amplification is reinforced by the networks' disordered nature, but
  saturates for high densities of active units. Our predictions 
  are quantitatively consistent with experiments on
  reconstituted tissues and actomyosin networks, and shed light on the role of the network
  microstructure in shaping active stresses in cells and tissue.
\end{abstract}

\maketitle

\begin{figure}[b!]
\includegraphics{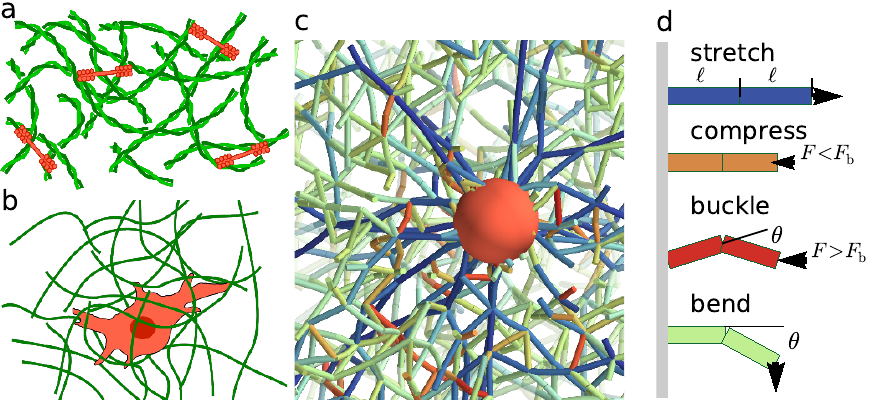}
\caption{\label{fig:3D}Biological fiber networks (green) transmit
  forces generated by localized active units (red).  \textbf{a.}~Myosin
  molecular motors exert forces on the actin
  cytoskeleton. \textbf{b.}~Contractile cells exert forces on the
  extracellular matrix. \textbf{c.}~The large nonlinear deformations induced by a model active unit in the surrounding fiber network result in stress amplification, as shown in this paper. See fiber color code
  in the next panel.  \textbf{d.}~Each bond in the network comprises two rigid
  segments hinged together to allow bending and buckling.}
\end{figure}

Living systems constantly convert biochemical energy into
forces and motion. In cells, forces are largely generated internally
by molecular motors acting on the cytoskeleton, a scaffold of protein
fibers (Fig.~\ref{fig:3D}\textbf{a}).  Forces from multiple motors are
propagated along this fiber network, driving numerous processes such
as mitosis and cell motility~\cite{Blanchoin:2014,Fletcher:2010}, and
allowing the cell as a whole to exert stresses on its surroundings. At
the larger scale of connective tissue, many such stress-exerting cells
act on another type of fiber network known as the extracellular
matrix (Fig.~\ref{fig:3D}\textbf{b}). This
network propagates cellular forces to the scale of the whole tissue,
powering processes such as wound healing~\cite{Ehrlich:1988} and
morphogenesis~\cite{Heisenberg:2013}.  Despite important differences
in molecular details and length scales, a common physical principle thus
governs stress generation in biological matter: internal forces from
multiple localized ``active units''---motors or cells---are propagated
by a fiber network to generate large-scale stresses. However, a
theoretical framework relating microscopic internal active forces to macroscopic
stresses in these networks is lacking.

This generic stress generation problem is confounded by the interplay
of network disorder and nonlinear elasticity.  Active units generate
forces at the scale of the network mesh size, and force transmission
to larger scales thus sensitively depends on local network
heterogeneities.  In the special case of linear elastic networks, the
macroscopic active stress is simply given by the density of active
force dipoles, irrespective of network
characteristics~\cite{Ronceray:2015}.  Importantly however,
this relationship is not applicable to most biological systems, since
typical active forces are amply sufficient to probe the nonlinear
properties of their constitutive fibers, which stiffen under tension and buckle
under compression~\cite{Broedersz:2014}. Indeed, recent experiments on
reconstituted biopolymer gels have shown that individual active units
induce widespread buckling and
stiffening~\cite{Lam:2011,Silva:2011,Murrell:2012}, and theory
suggests that such fiber nonlinearities can enhance the range of force
propagation~\cite{Shokef:2012,Notbohm:2014}.

Fiber networks also exhibit complex, nonlinear mechanical properties
arising at larger scales, owing to collective deformations favored by
the networks' weak connectivity~\cite{Broedersz:2014,Onck:2005,Heussinger:2007b,Conti:2009}. The role of connectivity in elasticity was
famously investigated by Maxwell~\cite{Maxwell:1864}, who noticed that
a spring network in dimension $d$ becomes mechanically unstable for
connectivities $z < 2d$. Interestingly, most biological fiber networks
exhibit connectivities well below this threshold, and therefore cannot
be stabilized solely by the longitudinal stretching rigidity of their
fibers. Instead, their macroscopic mechanical properties are
typically controlled by the fiber bending
rigidity~\cite{Broedersz:2011}. In contrast to stretching-dominated
networks with connectivities above the Maxwell threshold, such weakly
connected, bending-dominated networks are soft and extremely sensitive
to mechanical perturbations~\cite{Broedersz:2011,Ulrich:2013,Wyart:2008,Sheinman:2012a}. In these networks, stresses generated by
active units propagate along intricate force
chains~\cite{Heussinger:2007,Head:2005aa} whose effects on force transmission
remain unexplored.  Collections of such active units generate large
stresses, with dramatic effects such as macroscopic network
stiffening~\cite{Koenderink:2009,Jansen:2013,Broedersz:2011a} and network remodelling~\cite{Silva:2011,Murrell:2012,Bendix:2008,Kohler:2011,Carvalho:2013}.

Here we study the theoretical principles underlying stress generation
by localized active units embedded in disordered fiber networks
(Fig.~\ref{fig:3D}c).  We find that arbitrary local force
distributions generically induce large isotropic, contractile stress
fields at the network level, provided that the active forces are large
enough to induce buckling in the network.  In this case, the stress
generated in a biopolymer network dramatically exceeds the stress
level that would be produced in a linear elastic medium~\cite{Ronceray:2015,Ranft:2010}, implying a
striking network-induced amplification of active stress. Our findings
elucidate the origins and magnitude of stress amplification observed in experiments
on reconstituted tissues~\cite{Lam:2011,Jen:1982} and actomyosin
networks~\cite{Koenderink:2009,Carvalho:2013,Lemiere:2015}. We thus
provide a new conceptual framework for stress generation in biological
fiber networks.

\section*{A lattice model for elastic fiber networks}
\label{sec:model} 
We investigate force transmission using a lattice-based
fiber network model~\cite{Broedersz:2011,Das:2007aa}. In our model,
straight fibers are connected at each lattice vertex by crosslinks
that do not constrain their relative angles. Each lattice edge
represents a ``bond'' made of two straight segments and can thus
stretch, bend, or buckle (Fig.~\ref{fig:3D}d). Segments have
stretching rigidity $\mu$ and a rest length equal to one, implying a
stretching energy $\mu (\ell-1)^2 / 2$ per segment of length
$\ell$. The fiber bending rigidity is set to unity by penalizing
angular deflections $\theta$ between consecutive segments through a
bending energy $2\sin^2(\theta/2)$. Consequently, individual bonds
buckle under a critical force $F_{b}=1$, and we consider nearly
inextensible fibers by assuming $\mu\gg 1$ (henceforth we use
$\mu=10^3$).

Network disorder is introduced through bond depletion, \emph{i.e.}, by
randomly decimating the lattice so that two neighboring vertices are
connected by a bond with probability $p$. This probability controls
the network's connectivity, giving rise to distinct elastic regimes
delimited by two thresholds $p_{\rm cf}$ and $p_{\rm b}$. The network
is stretching-dominated for $p>p_{\rm cf}$, bending-dominated for
$p_{\rm b}<p<p_{\rm cf}$, and mechanically unstable for $p<p_{\rm
  b}$. Here we consider 2D hexagonal lattices, for which $p_{\rm
  b}\simeq 0.45$ and $p_{\rm cf}\simeq 0.65$, and 3D FCC lattices with
$p_{\rm b}\simeq 0.27$ and $p_{\rm cf}\simeq 0.47$. Since the network
displays singular behavior in the vicinity of $p_{\rm b}$ and $p_{\rm
  cf}$, here we focus our investigations on the generic stretching-
and bending-dominated regimes away from these critical
points~\cite{Broedersz:2011}.

We model active units as sets of forces $\mathbf{F}_i$ exerted on
network vertices $i$ with positions $\mathbf{R}_i$, and consider
networks at mechanical equilibrium under the influence of these
forces. We denote by $\sigma$ the trace (\emph{i.e.}, the isotropic
component) of the coarse-grained active stress induced in the fiber
network by a density $\rho$ of such units.

The relationship between this active stress and local forces in homogeneous linear networks is very simple, and yields~\cite{Ronceray:2015}
\begin{equation}
\label{eq:stress_lin}
\sigma=\sigma_\mathrm{lin} = -\rho \mathcal{D}_\mathrm{loc},
\end{equation}
where $\mathcal{D}_\mathrm{loc} = \sum_{i} \mathbf{F}_i \cdot
\mathbf{R}_i$ is the dipole moment of the forces associated with a
single active unit. Equation~(\ref{eq:stress_lin}) is generically
violated in disordered or nonlinear networks, although it holds on
average in linear networks with homogeneous disorder:
\begin{equation}\label{eq:disorderedSigma}
  \langle \sigma\rangle=\sigma_\mathrm{lin}, 
\end{equation}
where $\langle\cdot\rangle$ denotes the average over
disorder~\cite{Ronceray:2015}. To quantify violations of
Eq.~(\ref{eq:stress_lin}), we define the far-field force
dipole $\mathcal{D}_\mathrm{far}$ through the relation
\begin{equation}\label{eq:Deff}
\sigma = -\rho\mathcal{D}_\text{far}\quad \Rightarrow \quad \frac{\mathcal{D}_\mathrm{far}}{\mathcal{D}_\mathrm{loc}} = \frac{\sigma}{\sigma_\mathrm{lin}}.
\end{equation}
Conceptually, this far-field dipole characterizes the apparent
strength of an individual active unit renormalized by force transmission
in the disordered, nonlinear network.  It quantifies how
contractile (${\cal D}_\mathrm{far}<0$) or extensile (${\cal
  D}_\mathrm{far}>0$) the active medium is, and the dipole
amplification ratio
$\mathcal{D}_\mathrm{far}/\mathcal{D}_\mathrm{loc}$ (or equivalently the stress amplification ratio ${\sigma}/\sigma_\mathrm{lin}$) measures the
deviation from linear homogeneous force transmission.

\section*{Contractility robustly emerges from large local forces }
\label{sec:local}

\begin{figure*}
  \includegraphics{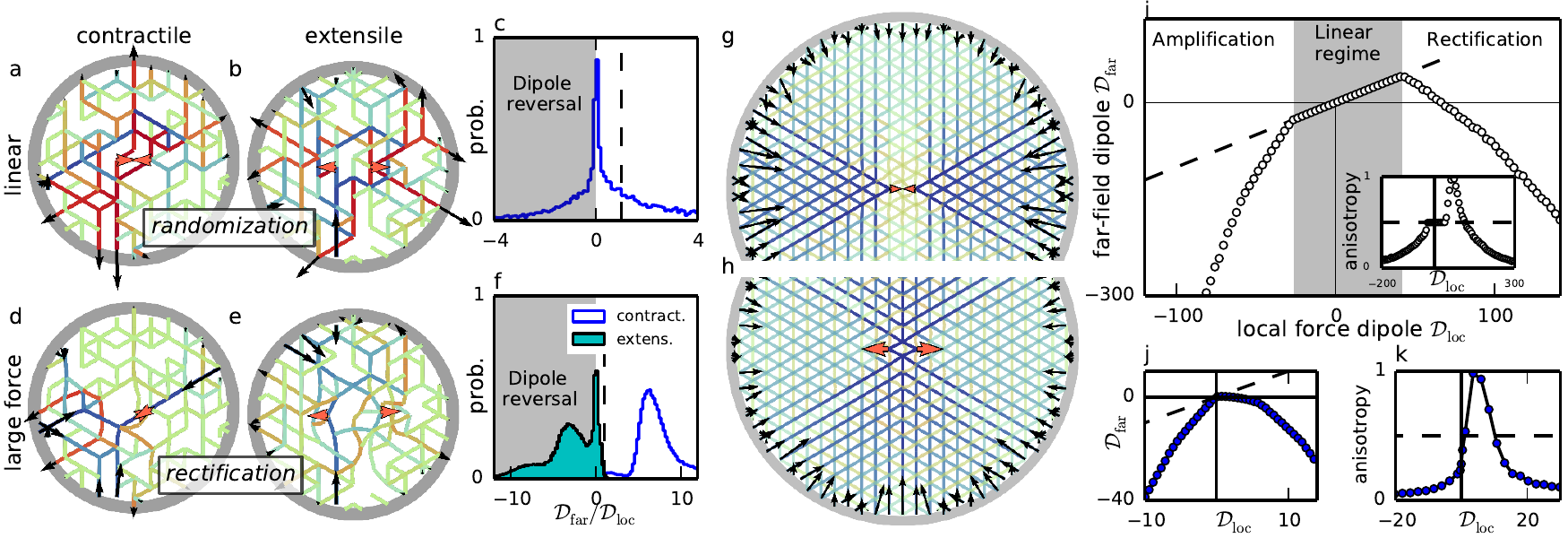}
  \caption{ \label{fig:local}Network buckling converts active forces
    into emergent isotropic contraction over a few mesh
    sizes. \textbf{a-b:} In the linear response regime ($F_0\ll F_b$),
    contractile and extensile active forces (red arrowheads) propagate
    along a complex network of force lines (blue=tension,
    red=compression), resulting in randomized force distributions at a
    fixed boundary (black arrows). \textbf{c:} The resulting
    distribution of dipole amplification ratios
    $\mathcal{D}_\mathrm{far}/\mathcal{D}_\mathrm{loc}$ is broad with
    widespread negative amplification (gray area) and an average equal
    to one (dashed line) ($n=10^4$ samples).  \textbf{d-e:} At larger forces
    (here $F_0=20F_b$), both contractile and extensile dipoles
    typically result in contractile forces at the
    boundary. \textbf{f:} Accordingly, the corresponding distribution
    of amplifications displays overwhelming negative amplification for
    locally extensile dipoles ($n=10^4$ samples). \textbf{g-h:} Regular
    networks subjected to large ($F_0=500F_b$), local dipoles of
    either sign exert uniformly contractile forces on the fixed
    boundary. \textbf{i:} Corresponding far-field dipole as a function
    of the local dipole, showing amplification and rectification in
    the nonlinear regime. \emph{Inset:} stress anisotropy parameter $1
    - (\sum_{\mu}\sigma^{\mu\mu})^2/(d \sum_{\mu,\nu}\sigma^{\mu\nu} \sigma^{\nu\mu} )$, as a
    function of the local dipole. Here $\sigma^{\mu\nu}$ is the
    coarse-grained active stress tensor of the active medium (see
    Supporting Information). \textbf{j-k:} Far-field dipole and
    anisotropy as a function of the local dipole in a
    bending-dominated 2D network ($p=0.6$, averaged over $n=10^4$
    samples). }
\end{figure*}

Stress generation by active units integrates mechanical contributions
from a range of length scales. We first consider the immediate
vicinity of the active unit. Network disorder plays a crucial role at
that scale, since forces are transmitted through a random pattern of
force lines determined by the specific configuration of depleted bonds
(Fig.~\ref{fig:local}\textbf{a-b}). To understand how these patterns
affect force transmission, we investigate the probability distribution
of the far-field force dipole $\mathcal{D}_\mathrm{far}$ for simple
active units consisting of two equal and opposite point forces of
magnitude $F_0$.

We first consider the linear regime $F_0\ll F_b$, where the average
dipole amplification equals unity:
$\langle\mathcal{D}_\mathrm{far}/\mathcal{D}_\mathrm{loc}\rangle=1$
[see Eqs.~(\ref{eq:disorderedSigma}-\ref{eq:Deff})].  The fluctuations
around this average are strikingly broad, as shown in
Fig.~\ref{fig:local}\textbf{c}. For instance, a significant fraction
($37\%$) of all network geometries yield negative amplification,
\emph{i.e.}, an effective extensility in response to a contractile
dipole (Fig.~\ref{fig:local}\textbf{b}). Overall, the far-field
response in the linear regime is only loosely correlated to the
applied force dipole.

The situation is dramatically different in the large force regime
($F_0\gg F_b$), where fibers buckle and induce nonlinear network
response. This is illustrated by the distributions of dipole
amplifications in two opposite cases: a large contractile and a large
extensile force dipole (Fig.~\ref{fig:local}\textbf{d-f}). First,
locally extensile dipoles predominantly undergo negative
amplification, implying far-field contractility irrespective of the
sign of $\mathcal{D}_\mathrm{loc}$ (as in, \emph{e.g.},
Fig.~\ref{fig:local}\textbf{e}). Second, the randomization observed in
the linear regime is strongly attenuated, and the sign of the
amplification is very reproducible (positive for $98\%$ of the
contractile dipoles and negative for $86\%$ of the extensile
ones). Third, the magnitude of the average amplification is
significantly larger than one (in Fig.~\ref{fig:local}\textbf{f}
$\av{\mathcal{D}_\mathrm{far}/\mathcal{D}_\mathrm{loc}} = 6.9$ and
$-3.2$ for contractile and extensile dipoles, respectively).

To understand these three effects, we consider contractile and
extensile dipoles in a simpler regular network (no bond depletion,
Fig.~\ref{fig:local}\textbf{g-h}). Qualitatively, these uniform networks behave
similarly to the randomly depleted ones described above: force dipole
conservation holds for $F_0\ll F_b$, while for $F_0\gg F_b$ dipoles
are rectified towards contraction and their magnitude is amplified
(Fig.~\ref{fig:local}\textbf{i}). The origin of these behaviors is
apparent from the spatial arrangement of the forces in
Figs.~\ref{fig:local}\textbf{g-h}. While contractile and extensile
active units both induce compressive and tensile stresses in their
immediate surroundings, the buckling of the individual bonds prevents
the long-range propagation of the former. This results in enhanced
tensile stresses in the far-field, and thus in strongly contractile
far-field dipoles. In addition, this nonlinear response renders the
far-field stresses uniformly tensile, and therefore more isotropic
than the active unit forces. We quantify this effect in the inset of
Fig~\ref{fig:local}\textbf{i} using an anisotropy parameter for the
far-field stresses, which indeed becomes very small for both positive
or negative large local dipoles.

Moving to a systematic quantification of force transmission in
depleted, bending-dominated networks, we show in
Fig~\ref{fig:local}\textbf{j-k} the same three effects of
rectification, amplification and isotropization, which set in at
smaller forces than in regular networks.  Overall, these effects are
very general and hold in both bending- and stretching-dominated
depleted networks, in two and three dimensions, and for active units
with complicated force distributions (see supporting Figs.~S3 and~S4). Thus,
beyond the immediate neighborhood of the active force-generating unit,
strong isotropic contractile stresses emerge in the system from a
generic local force distribution due to the nonlinear force
propagation properties of the fiber network.

\begin{figure*}
  \includegraphics{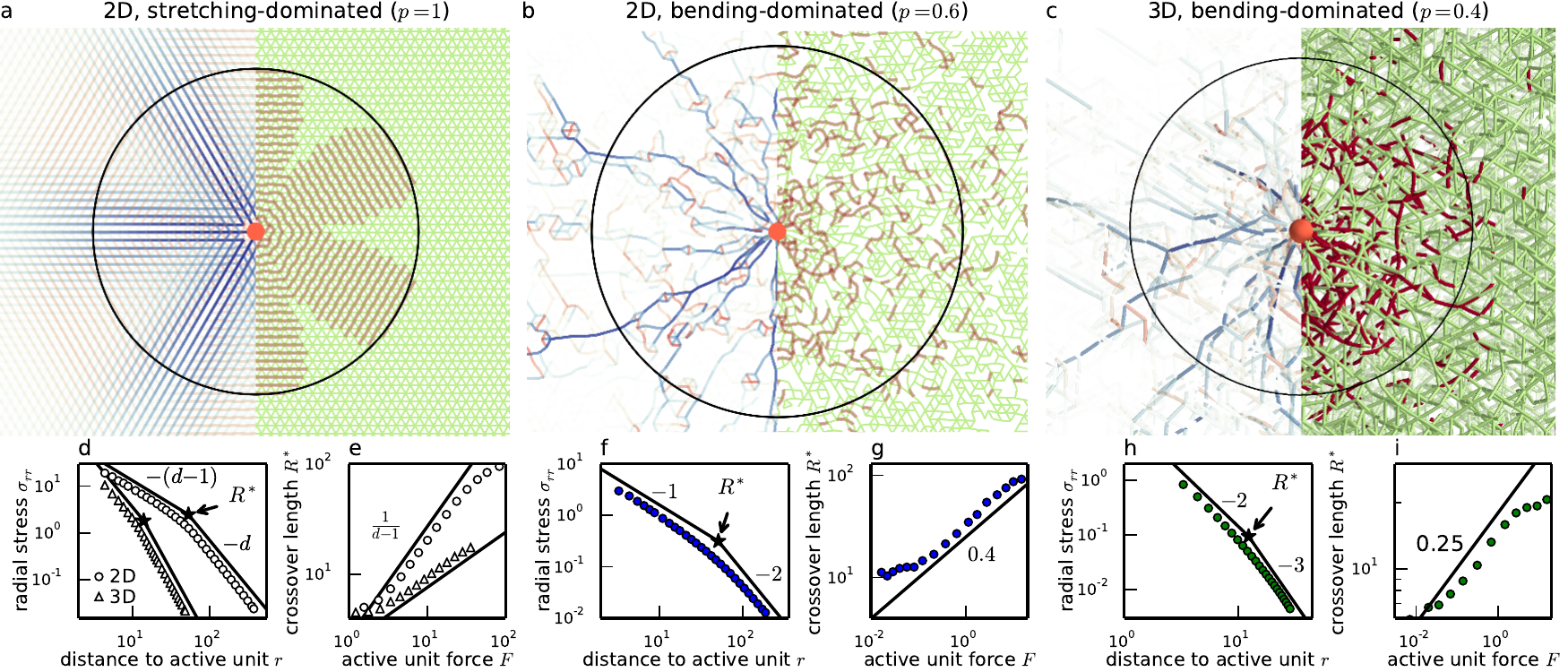}
  \caption{ \label{fig:single_motor} Nonlinear network behavior
    enhances the range over which stresses are transmitted.
    \textbf{a-c:} A localized, isotropically pulling active unit (red
    circle of radius $R_0=1.95$) induces stress lines (left side of
    each panel, blue=tension and red=compression) and buckling (right
    side of each panel, in red; non-buckled bonds are green) in the surrounding fiber
    network. Black circle: radius $R^*$ of the rope-like region. Panel
    \textbf{c} shows a slice of a 3D system. \textbf{d, f} and
    \textbf{h:} Decay of the average radial stress in the network
    (corrected for boundary effects, see Supporting Information) as a
    function of the distance to the active unit. Fitting the curve
    with the power laws of Eqs.~(\ref{eq:rope}-\ref{eq:linforceprop})
    yields a measure of the crossover radius $R^*$. \textbf{e, g} and
    \textbf{i.}  We tentatively describe the dependence of the crossover radius on
      active unit force by a power law (solid line) in the intermediate-$F$ regime where
      it is not complicated by finite size effects due to either the active unit size (at small $F$)
      or that of the system
      boundary (at large $F$).  Results obtained in a 2D circular (3D spherical) network
    of radius $200$ ($33$) with fixed boundaries and averaged over
    $100$ samples for disordered networks.}
\end{figure*}

\section*{A model for active units as isotropic pullers}
While nonlinear force transmission over large length scales involves
large active forces, the model for active units used above can only
exert moderate dipoles in soft, weakly connected networks. Indeed, for
large enough contractile dipoles the two vertices on which the forces
are applied collapse to a point (Fig.~\ref{fig:local}\textbf{d}),
preventing further contraction. In contrast, molecular motors and
contractile cells continuously pull fibers in without collapsing. To
reflect this, we introduce an active unit capable of exerting
arbitrarily large forces without changing its size. The unit is
centered on a vertex $i$, and pulls on any vertex $j$ within a
distance $2R_0$ with a radial force
\begin{equation} \label{eq:puller}
\mathbf{F}_{ij}=
\begin{cases}
-F_0 \frac{{r}_{ij}}{R_0}\mathbf{\hat{r}}_{ij}
&\mbox{if } r_{ij}<R_0\\
-F_0(2-\frac{{r}_{ij}}{R_0})\mathbf{\hat{r}}_{ij}
&\mbox{if } R_0\leqslant r_{ij}<2R_0\\
\end{cases},
\end{equation}
where $F_0$ is the maximum force exerted by the unit on a vertex,
$r_{ij}$ is the distance between $i$ and $j$ and
$\mathbf{\hat{r}}_{ij}$ is the associated unit vector.  A strong
active unit in a soft network may pull in many fibers, exerting a
force $\approx F_0$ on each of them. Adding the contributions of all
these fibers results in a large local dipole, the magnitude of which
is not well reflected by the value of $F_0$. The influence of the
active unit on the surrounding network is better characterized by the
force $F$, which we define as the average force per unit area exerted
on the surrounding network by the active unit at its outer surface
($r=2R_0$). Finally, we assign an isotropic force distribution to the
active puller defined in Eq.~(\ref{eq:puller}). This choice is
justified by the observation that anisotropic force
distributions are rectified towards isotropy by the network
(Figs.~\ref{fig:local}\textbf{i, k}).

\section*{Contractile forces are long-ranged in bucklable media}

We now study force propagation beyond the immediate vicinity of an
active unit (Fig.~\ref{fig:single_motor}) using the above-described
isotropic puller.  Simple theoretical arguments dictate two asymptotic
regimes for this propagation. Close to the active unit, forces are
large and fiber buckling affects force transmission, while beyond a
crossover distance $R^*$ forces are weak and linear elasticity
prevails.

To describe the near-field regime, we note that fiber buckling
prevents the network from sustaining compressive stresses above the
buckling threshold. Close to the active unit, the network is thus
effectively equivalent to a network of floppy ropes. The active unit
pulls on these ropes, and thus becomes the center of a radial
arrangement of tensed ropes. Force balance on a small portion of a
spherical shell centered on the active unit imposes that radial
stresses in this rope-like medium decay as
\begin{equation}
\label{eq:rope}
\sigma_{rr}(r) \propto r^{-(d-1)},\quad r < R^*
\end{equation}
where $r$ is the distance from the active unit and $d$ the dimension
of space~\cite{Rosakis:2014}. In the far field, stresses are small and
buckling does not occur, implying that force transmission crosses over
from rope-like to linear elasticity:
\begin{equation}
 \label{eq:linforceprop}
 \sigma_{rr}(r) \propto r^{-d}, \quad r > R^*.
\end{equation}
Stress decay is thus significantly slower in the rope-like near-field
than in the linear far-field, leading to an increased range for force
transmission~\cite{Notbohm:2014}.  Conceptually, the faster decay in a
linear elastic medium can again be understood by balancing forces on a
fraction of spherical shell centered on the active unit, where radial
stresses are now partially compensated by orthoradial stresses. We
expect that the crossover between these two regimes occurs when radial
stresses are comparable to the buckling stress, implying that the
crossover length depends on the active force:
\begin{equation}\label{eq:Rstar}
R^* \approx R_0 \left(\frac{F}{F_{\rm b}}\right)^{1/(d-1)}
\end{equation}

To test this two-regime scenario, we simulate force propagation away
from a single active unit in both stretching- and bending-dominated
networks in two and three dimensions.  In all cases, rope-like radial
stresses and bond buckling are predominant in the vicinity of the
active unit (Fig.~\ref{fig:single_motor}\textbf{a-c}). Monitoring the
decay of radial stresses with $r$, we find an apparent crossover from
rope-like to linear behavior, consistent with
Eqs.~(\ref{eq:rope}) and (\ref{eq:linforceprop})
(Fig.~\ref{fig:single_motor}\textbf{d, f, h}).  

Visually, the crossover length $R^*$ coincides with the outer boundary
of the radially tensed, buckling-rich region
(Fig.~\ref{fig:single_motor}\textbf{a-c}, black circles).  In
stretching-dominated networks, our prediction of Eq.~(\ref{eq:Rstar})
captures the force dependence of this crossover length
(Fig.~\ref{fig:single_motor}\textbf{e} and S5). In contrast,
bending-dominated networks display a more complex behavior: while the
system still exhibits a transition from rope-like to linear force
transmission, the crossover region is much broader
(Fig.~\ref{fig:single_motor}\textbf{f, h}) and forces propagate along
heterogeneous patterns reminiscent of previously reported force chains
(Fig.~\ref{fig:single_motor}\textbf{b-c})~\cite{Heussinger:2007,Dasanayake:2011}.
This strong concentration of the tensile stresses allows rope-like
force transmission to extend much further than predicted by
Eq.~(\ref{eq:Rstar}).  Instead, we find behavior that is reasonably
well described by a power law $R^*\propto F^\alpha$ with anomalous
exponents $\alpha\approx 0.4$ in 2D and $\alpha\approx 0.25$ in 3D
(Fig.~\ref{fig:single_motor}\textbf{g, i}). These exponents appear to
be insensitive to the exact value of the depletion parameter $p$
within the bending-dominated regime (Supporting Fig.~S5). The
difference between stretching- and bending-dominated exponents
suggests elastic heterogeneities qualitatively affect force
transmission in such soft networks.  As a result, contractile forces
large enough to induce buckling benefit from an enhanced range of
transmission, characterized by the mesoscopic radius of the rope-like
region $R^*$.

\section*{Amplification by a collection of active units}

\begin{figure*}
  \includegraphics{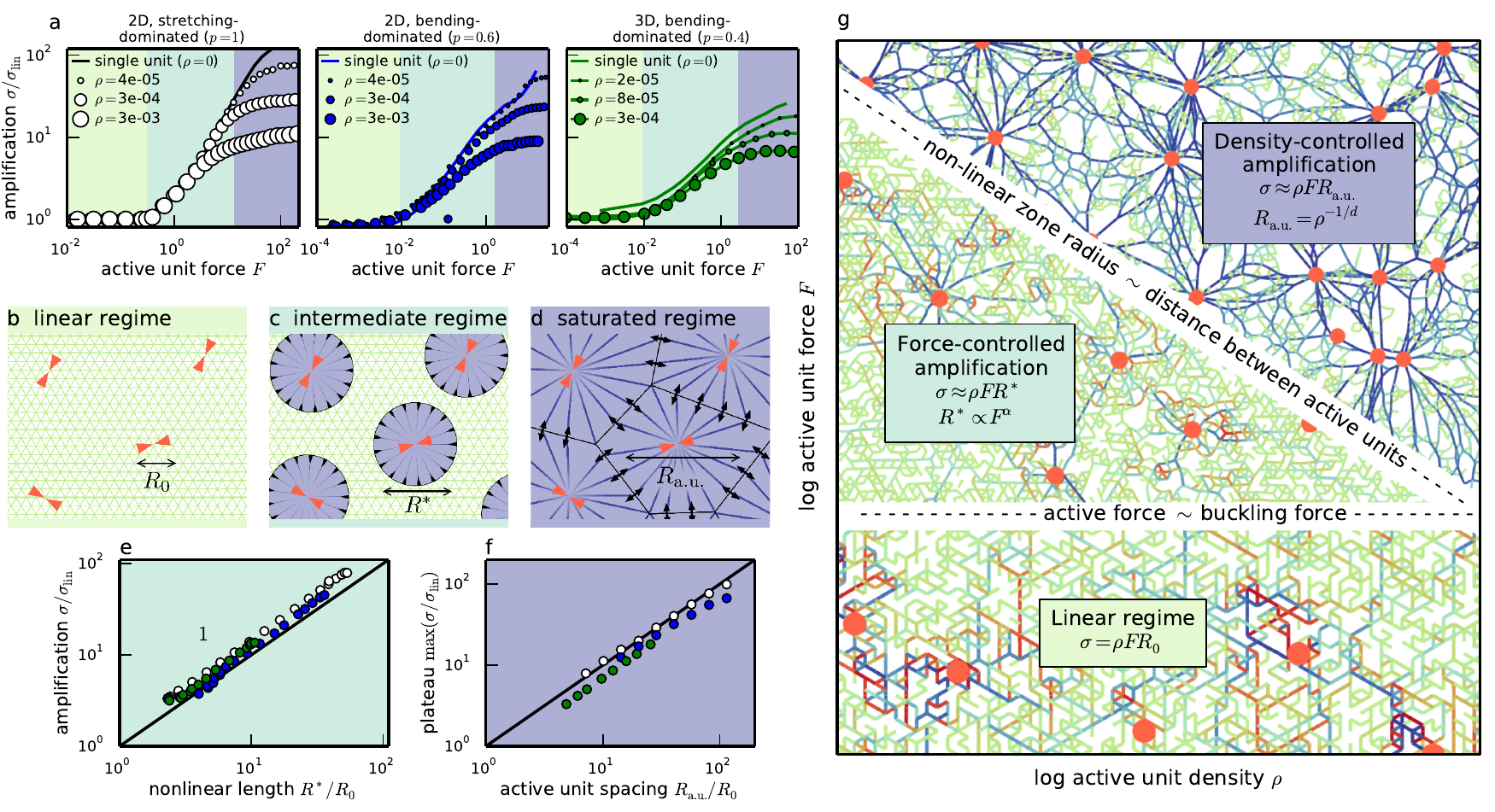}
  \caption{ \label{fig:multimotors} Force transmission in the presence
    of a finite density $\rho$ of active units.  \textbf{a.}~Fiber
    networks in different dimensions and elastic regimes all display
    three stress amplification regimes as a function of active unit
    density and force, as suggested by the colored background. \textbf{b-d.}~Schematics of the network structure in
    each regime. The low-force linear regime (\textbf{b}.) transitions to a regime
    of nonoverlapping nonlinear regions (\textbf{c}.) as soon as $F$ is sufficient
    to induce buckling. These nonlinear regions grow with increasing
    $F$, and amplification saturates as they start overlapping, which
    turns the whole network into a rope network (\textbf{d}.).  \textbf{e.}~In the
    intermediate force regime, the stress amplification ratio is equal
    to the ratio $R^*/R_0$ as predicted by
    Eq.~(\ref{eq:active_stress_growth}).  \textbf{f.} In the
    large-force regime, the stress amplification ratio is equal to the
    ratio $R_\mathrm{a.u.}/R_0$ as predicted by
    Eq.~(\ref{eq:stress-sublinear}).  \textbf{g.}~Schematic phase
    diagram indicating the domain of applicability of the three stress
    amplification regimes, with representative snapshots of the
    corresponding systems in the background.}
\end{figure*}

Over large length scales, active stresses in biological systems are
generated by multiple active units.  We thus compute the stress
amplification ratio in the presence of a finite density of randomly
positioned active units in 2D and 3D for various densities $\rho$ and
depletion parameters $p$ (Fig.~\ref{fig:multimotors}\textbf{a}). In all cases we observe three stress
amplification regimes as a function of the unit force $F$: a low-force
plateau without amplification, an intermediate regime of increasing
amplification and a saturation of the amplification at a level that
depends on $\rho$.

In the low-force regime, linear force transmission prevails
(Fig.~\ref{fig:multimotors}\textbf{b}) and the active stress is given
by Eq.~(\ref{eq:stress_lin}):
\begin{equation}
  \sigma = \sigma_\mathrm{lin} =  -\rho \mathcal{D}_\mathrm{loc} \approx \rho F R_0 \label{eq:linear_stress}.
\end{equation}

For moderate forces, the fibers in the network buckle in the vicinity
of each active unit, up to a distance $R^*$. Individual units are thus
typically surrounded by \emph{nonoverlapping} nonlinear regions of
size $R^*$, as illustrated in Fig.~\ref{fig:multimotors}\textbf{c}. To
predict the resulting active stress in the system, we model each
nonlinear region as an effective active unit of size $R^*$ and force
dipole $\mathcal{D}_\mathrm{eff}\approx -FR^*$, where we used
Eq.~(\ref{eq:rope}) to describe force propagation within the nonlinear
region. As the effective units are themselves embedded in a linear
medium, linear force transmission [Eq.~(\ref{eq:stress_lin})] outside
of these units implies
\begin{equation}
  \label{eq:active_stress_growth}
  \sigma \approx -\rho \mathcal{D}_\mathrm{eff} \approx \rho F  R^*.
\end{equation}
We thus predict that stress amplification in this regime scales as
$\sigma/\sigma_\mathrm{lin} \approx R^*/R_0$.  We confirm this
prediction in Fig.~\ref{fig:multimotors}\textbf{e}. Since $R^*$
increases with the active unit force in this regime, the large-scale
stress amplification $\sigma/\sigma_\mathrm{lin}$ increases with $F$ as previously observed in Fig.~\ref{fig:multimotors}\textbf{a}.

For large forces, the radius of the rope-like regions becomes so large
as to exceed the typical distance between adjacent active units
$R_\mathrm{a.u.}= \rho^{-1/d}$. This causes the nonlinear regions
associated to neighboring active units to overlap, and renders the
whole network mechanically equivalent to a collection of tensed,
inextensible ropes whose geometry does not change significantly upon
further increase of the force.  To estimate the resulting network
stress, we approximate the system as a mosaic of effective active
units of size $R_\mathrm{a.u.}$ each with a force dipole
$\mathcal{D}_\mathrm{eff}\approx -FR_\mathrm{a.u.}$
(Fig.~\ref{fig:multimotors}\textbf{d}). This yields
\begin{equation}
  \sigma \approx -\rho \mathcal{D}_\mathrm{eff} \approx \rho F R_\mathrm{a.u.} =  \rho^{1-1/d}   F.
   \label{eq:stress-sublinear}
\end{equation}
The resulting prediction for the stress amplification,
$\sigma/\sigma_\mathrm{lin} \approx R_\mathrm{a.u.}/R_0$, is confirmed
in Fig.~\ref{fig:multimotors}\textbf{f}. Strikingly, the stress
generated in this large-force regime has a nonlinear dependence on
$\rho$, again consistent with Fig.~\ref{fig:multimotors}\textbf{a}. Indeed, the addition or removal of active units leads to large
rearrangements of the rope network, resulting in significant local
modifications of force transmission.

We summarize the physics of collective stress-generation by many
active units in a phase diagram (Fig.~\ref{fig:multimotors}\textbf{g}). In each regime, the magnitude of an active unit's effective force dipole is directly proportional to one of the three length scales $R_0$, $R^*$ and $R_\mathrm{a.u.}$ [Eqs.~(\ref{eq:linear_stress}-\ref{eq:stress-sublinear})]. While
we have shown that $R^*$ depends on the dimensionality and
connectivity of the network, the other two length scales are purely
geometrical. An important consequence of these findings is that the
active stress generated in the associated regimes is essentially
independent of the detailed properties of the fiber network.

\section*{Discussion}
In living organisms, microscopic units exert active forces that are
transmitted by fibrous networks to generate large-scale stresses. The
challenge in analyzing this force-transmission problem stems from the
disordered architecture of such fibrous networks and the
nonlinearities associated with the strong forces exerted by biological
active units. Despite this complexity, we find surprisingly simple and
robust behaviors: in response to any distribution of active forces,
dramatically amplified contractile stresses emerge in the network on
large scales.  This remarkable property hinges only on the local
asymmetry in elastic response between tensed and compressed fibers,
and is enhanced by network disorder.  Our simple, yet realistic
description of individual fibers yields a universal scenario for force
transmission: long-ranged, rope-like propagation near a strong active
unit, and linear transmission in the far-field. This generic result should be
contrasted with recent studies focused on fibers with special singular force-extension relation~\cite{Notbohm:2014,Rosakis:2014} and resulting in non-universal force transmission regimes.

\begin{table*}
  \caption{\label{tab:table} Experimental data
    support stress amplification in fiber networks. The rope-like
    radius $R^*$, linear-theory active stress $\sigma_\textrm{lin}$
    and predicted amplified stress $\sigma_\textrm{th}$ are computed
    using Eqs.~(\ref{eq:linear_stress}-\ref{eq:stress-sublinear}) from
    independent estimates of the single-unit force $F$ (see Supporting
    Information) for comparison with the experimentally measured
    active stress $\sigma_\textrm{exp}$. We use the
    stretching-dominated scaling for $R^*$ [Eq.~(\ref{eq:Rstar})], and
    thus the predicted active stress in system~II is a lower bound as
    indicated by the ``$\geqslant$'' symbol in the
    $\sigma_\textrm{th}$ column; the ``$\geqslant$'' in the
    $\sigma_\textrm{exp}$ column reflects experimental uncertainties.}
\begin{tabular}{rl c c c c c c }
& System & $\quad R_0\quad$ & $\quad R_\textrm{au}\quad$ & $\qquad R^*=R_0(F/F_b)^{1/(d-1)}\qquad$ & $\quad\sigma_\textrm{lin}\quad$ & $\qquad\qquad\sigma_\textrm{th}\qquad\qquad$ & $\quad\sigma_\textrm{exp}\quad$ \\
\hline
I& 3D actomyosin~\cite{Koenderink:2009} &
$1\,\mu\rm{m}$ & $1\,\mu\rm{m}$ & $0.3\,\mu\rm{m}$ (linear regime) &
$12\,\rm{Pa}$ & $12\,\rm{Pa}$ & $14\,\rm{Pa}$\\

II& 2D actomyosin~\cite{Carvalho:2013,Lemiere:2015} &
$1\,\mu\rm{m}$ & $20\,\mu\rm{m}$ & $15\,\mu\rm{m}$ (force-controlled) &
$0.014\,\rm{pN}/\mu\rm{m}$ & $\geqslant 0.2\,\rm{pN}/\mu\rm{m}$ & $\geqslant 1\,\rm{pN}/\mu\rm{m}$\\

III& 3D blood clot~\cite{Lam:2011,Jen:1982} &
$2\,\mu\rm{m}$ & $15\,\mu\rm{m}$ & $70\,\mu\rm{m}$ (density-controlled) &
$9\,\rm{Pa}$ & $70\,\rm{Pa}$ & $150\,\rm{Pa}$\\
\end{tabular}
\end{table*}

Our generic phase diagram (Fig~\ref{fig:multimotors}\textbf{g})
recapitulates our quantitative understanding of stress generation by a
collection of active units based on the interplay between three length
scales: active unit size $R_0$ , rope-like length $R^*$, and typical
distance between units $R_\mathrm{a.u.}$. To validate these
predictions, we compare them with existing measurements on a broad
range of \emph{in vitro} systems (Table~\ref{tab:table}). We first
consider system~I, a dense three-dimensional actin network with mesh
size $\simeq 200\,\textrm{nm}$ in the presence of myosin motors, which
assemble into so-called myosin thick filaments. A thick
filament---which we consider as an individual active unit---exerts a
typical force $F=6\,\mathrm{pN}$, much smaller than the buckling
threshold $F_b\approx 50\,\mu\mathrm{m}$ associated with a single
$200\,\textrm{nm}$-bond. This implies an active stress identical to
the linear prediction, as confirmed by the experimental
result~\cite{Koenderink:2009}. We next consider system~II, a
two-dimensional actin network bound to the outer surface of a lipid
vesicle. The active units are essentially the same as in System~I, but
are much more sparsely distributed ($R_\mathrm{a.u.}\simeq
20\,\mu\textrm{m}$). The network in system~II is also much looser
($\textrm{mesh size}\simeq 1\,\mu\textrm{m}$) than in system~I,
resulting in a much smaller bond buckling force. The combination of a
low buckling threshold and a large spacing between active units leads
us to predict a significant stress amplification $R^*/R_0\simeq 15$
associated to the force-controlled regime
(Fig~\ref{fig:multimotors}\textbf{c, g}), in reasonable agreement with
experiments~\cite{Carvalho:2013,Lemiere:2015}.  Finally, we consider a
clot comprised of fibrin filaments and contractile platelets as active
units (system~III).  The large forces exerted by platelets allow for
long-range nonlinear effects, placing this \emph{in vitro} system deep
in the density-controlled regime
(Figs.~\ref{fig:multimotors}\textbf{d, g}). Consequently, we expect
stress amplification to be controlled by the distance between active
units, irrespective of the large value of the active force $F\approx
10^5\,F_b$. We thus predict an amplification factor
$R_\mathrm{a.u}/R_0 \simeq 8$, in good agreement with experimental
data~\cite{Lam:2011,Jen:1982}. These three examples demonstrate our
theory's ability to quantitatively account for stress amplification,
and recent progress in the micromechanical characterization of active
fiber networks opens promising perspectives for further exploring
active stress
amplification~\cite{Lam:2011,Murrell:2012,Carvalho:2013}.

Far from merely transmitting active forces, we show that fiber
networks dramatically alter force propagation as contractility emerges
from arbitrary spatial distributions of local active forces. This
could imply that living organisms do not have to fine-tune the
detailed geometry of their active units, since any local force
distribution yields essentially the same effects on large length
scales.  This emergence of contractility sheds a new light on the
longstanding debate in cytoskeletal mechanics regarding the emergence
of macroscopic contraction in non-muscle actomyosin despite the
absence of an intrinsic contractility of individual myosin
motors~\cite{Dasanayake:2011,Hatano:1994,Sekimoto:1998,Lenz:2012,Mizuno:2007aa,Murrell:2015}.
Indeed, while these motors exert equal amounts of local pushing and
pulling forces~\cite{Lenz:2012a,Lenz:2014}, our result suggests that
the surrounding network rectifies pushing contributions into uniform
contraction. More broadly, we suggest that this strong propensity for
the emergence of contraction in fibrous materials can explain the
overwhelming dominance of contractile stresses in active biological
materials up to the tissue level. Clearly, this does not mean that it
is impossible to generate large-scale expansion in living organisms as
required for limb abduction and extension or for lung
inflation. Nevertheless, in each of these examples the expansion
actually results from the clever harnessing of muscle contraction
through lever structures involving our skeleton.

Our results suggest a novel design principle for active fiber networks
geared to maximize stress-generation. In a linear medium, the stress
generated does not depend on the spatial distribution of active
units. In contrast, we predict that in fiber networks, larger stresses
can be obtained by clustering the active units. Such regrouping of a
set number of force generators to enhance stress amplification could
play a role in smooth muscle, where the number of myosins in
individual thick filament is regulated
dynamically~\cite{Seow:2005}. Similarly, at the tissue level,
clustering of contractile cells occurs during wound
repair~\cite{Rocha-Azevedo:2013}.

Our findings connect widely used ``active gels'' phenomenological theories~\cite{Prost:2015} to their underlying molecular foundation, a crucial step in bringing theory and experiments together in the study of active biological matter, and calls for further progress in characterizing force transmission in more complex fiber networks. Finally, beyond biopolymer networks our work opens avenues to
understand force transmission in novel metamaterials whose macroscopic
properties crucially hinge on their microscopic
buckling~\cite{Kang:2013,Florijn:2014}.

\begin{acknowledgments}
  We thank C\'ecile Sykes and Guy Atlan for fruitful discussions. This
  work was supported by grants from Universit\'e Paris-Sud and CNRS,
  the University of Chicago FACCTS program, Marie Curie Integration
  Grant PCIG12-GA-2012-334053 and ``Investissements d'Avenir'' LabEx
  PALM (ANR-10-LABX-0039-PALM) to ML as well as the German Excellence
  Initiative via the program `NanoSystems Initiative Munich' (NIM) and
  the Deutsche Forschungsgemeinschaft (DFG) via project B12 within the
  SFB 1032.  PR is supported by ``Initiative Doctorale
  Interdisciplinaire 2013'' from IDEX Paris-Saclay, and CPB is
  supported by a Lewis-Sigler fellowship. ML's group belongs to the
  CNRS consortium CellTiss. Figures realized with
  Matplotlib~\cite{Hunter:2007} and Mayavi2~\cite{Ramachandran:2011}.
\end{acknowledgments}

\nocite{Ronceray:2015,Julicher:2007,Joanny:2009,Prost:2015,Koenderink:2009,Rosenfeld:2003,Lemiere:2015,Jen:1982,Lam:2011,Storm:2005}

\bibliographystyle{apsrev4-1}
\bibliography{Bibliography}

\end{document}